\documentclass[aps,preprint,floatfix,nofootinbib,showpacs]{revtex4-1}
\pdfoutput=1
\usepackage{graphicx,color}
\usepackage{hyperref}
\begin{document}

\title{Implications on the Heavy CP-even Higgs Boson\\
from Current Higgs Data}

\renewcommand{\thefootnote}{\arabic{footnote}}

\author{
Jung Chang$^1$, Kingman Cheung$^{1,2}$, Po-Yan Tseng$^1$, 
and Tzu-Chiang Yuan$^3$}
\affiliation{
$^1$ Department of Physics, National Tsing Hua University,
Hsinchu 300, Taiwan \\
$^2$ Division of Quantum Phases and Devices, School of Physics, 
Konkuk University, Seoul 143-701, Republic of Korea \\
$^3$ Institute of Physics, Academia Sinica, Nangang, Taipei 11529, Taiwan 
}
\date{\today}

\begin{abstract}
  The current Large Hadron Collider data indicates that the newly
  observed resonance has the $WW$ and $ZZ$ modes consistent with the
  standard model Higgs boson, while the $\gamma\gamma$ mode is about
  $1.2 - 2$ times that of the standard model prediction and the
  tau-pair mode is suppressed. If this pattern persists in the
  upcoming data, it would be a sign for physics beyond the standard
  model.  In the type II two-Higgs-doublet model, it is the region
  where $\sin\alpha \approx 0$ and a moderately large $\tan\beta = 10
  - 20$ that the lighter CP-even Higgs boson can accommodate the
  current data. We note that in this region the heavier CP-even Higgs
  boson must have a large decay branching ratio into tau pairs.
  We find that this heavier Higgs boson can be observable 
  in the associated production with a $b\bar b$ pair and through  
  the decay into a tau pair.
\end{abstract}

\maketitle

\section{Introduction}
It is of very high expectation that the observed particle at the Large
Hadron Collider (LHC) \cite{atlas,cms} is the long-sought Higgs boson 
of the standard model (SM),
which was proposed in 1960s \cite{higgs}.  At the end of 2011, both
the ATLAS and CMS \cite{cms-atlas} experiments at the 
LHC have seen some excess of events of a possible Higgs boson
candidate in the decay modes of $h\to \gamma\gamma$, $h\to WW^* \to \ell^+
\nu \ell^- \bar \nu$ and $h \to Z Z^* \to 4\ell$ channels.  Finally, the
discovery was announced in July 2012 by ATLAS \cite{atlas} and CMS
\cite{cms}. The channels $WW$ and $ZZ$ are consistent with the 
predictions of the SM Higgs boson, while the $\gamma\gamma$ rate is 
somewhat higher than expectation.  Some
evidence is seen in the $b\bar b$ mode at the Tevatron
\cite{tevatron}, but the mass range is quite wide.  On the other hand,
the $\tau^+\tau^-$ mode appears to shy away from the detectors, 
albeit the data contain large uncertainties.

The diphoton production rate is about a factor of $1.2-2$ as much as
that of the SM Higgs boson.
A large number of models have 
been put forward to account for the observed particle at 125 GeV,
including the SM, MSSM, NMSSM, $U(1)$-extended minimal supersymmetric 
standard model, and other MSSM-extended models,
fermiophobic Higgs, 2HDM of various types, Randall-Sundrum radion, 
inert-Higgs doublet, triplet Higgs models 
(a summary of various models can be found in Ref.~\cite{ours}.)
They all can explain the enhanced diphoton rate with some choices of 
parameter space.
Nevertheless, the most peculiar observation is that the $\tau^+\tau^-$ 
mode is suppressed \cite{private} in the data.
\footnote
{There was an update in November 2012: the $\tau\tau$ mode is 
measured to be $0.7 \pm 0.5$ (CMS) and $0.7 \pm 0.7$ (ATLAS) \cite{private}.
The $\tau\tau$ mode is still suppressed but with a large uncertainty.}
If this picture persists in the
upcoming data, it would be a sign for physics beyond the SM.

The two-Higgs-doublet model (2HDM) has enough free parameters
that allow one to achieve a light CP-even Higgs boson of 125 GeV, 
for which the $WW^*$ and $ZZ^*$ modes 
are consistent with the SM Higgs boson, the diphoton mode is enhanced,
and $\tau^+\tau^-$ mode is suppressed.  This can be achieved in the type II 2HDM
in the following parameter space
\begin{equation}
\label{parameters}
\sin\alpha \approx 0 \qquad {\rm and} \qquad \tan\beta = 10-20 \; .
\end{equation}
In this case, 
$\sin(\beta - \alpha) \approx \sin\beta$, which is close to 1 for 
large enough $\tan\beta$. Therefore, the $WW^*$ and $ZZ^*$ modes are about the
same as the SM, and the $\tau^+\tau^-$ mode is highly suppressed. With
a reduced total width the $\gamma\gamma$ branching ratio is enhanced.
In fact, this is the favorable region for enhancement in the diphoton 
mode obtained in Ref.~\cite{sher}. Other related works in 2HDM can be
found in Refs.~\cite{2hdm-other,2hdm-heavy}

In this Letter, we show that with the choice of the parameter space of 
Eq.(\ref{parameters})
the lighter CP-even Higgs boson $h$ is consistent with the observed data, 
and the heavier CP-even Higgs boson $H$ 
can be observable through the associated production with a $b\bar b$ pair
in the $b\bar b \tau^+ \tau^-$ final state at the 8 TeV and 14 TeV LHC.
This is the main result of the paper.

\section{Two-Higgs-Doublet Model}
Instead of one, the 2HDM employs two Higgs doublets.
In order to avoid dangerous tree-level flavor-changing neutral currents, 
a discrete $Z_2$ symmetry is usually imposed in all the popular 2HDMs. 
A number of possible arrangements of the two doublets and various scenarios 
of the mass spectra for the Higgs bosons are recently analyzed in 
Refs.~\cite{cheon,thomas,fox,kang,bai,droz} in light of current data.  
In this work, we focus on the type II, which has the same Higgs sector 
as the MSSM. 
\footnote
{
Though the type II model with the light CP-even Higgs $m_h\approx 125$ GeV 
is not the best fit to the current Higgs data,
it still provides a decent good fit \cite{kang,bai}.
}
The Higgs sector consists of two Higgs doublets
$H_u = \left( H_u^+ \, H^0_u \right)^T$ and $H_d = \left( H_d^+ \, H^0_d \right)^T$
where the subscripts $u,d$ denote the right-handed quark singlet fields
that the Higgs doublets couple to. 
After electroweak symmetry breaking (EWSB),
there are two CP-even, one CP-odd, and a pair of charged Higgs bosons.
The parameters of the model in the CP-conserving case can be chosen as
\[
  m_h,\; m_H,\; m_A, \; m_{H^+},\; \tan\beta \equiv \frac{v_u}{v_d},\;
 \alpha
\]
where $\alpha$ is the mixing angle between the two CP-even Higgs bosons.
All these are free parameters of the model but subjected 
to theoretical and experimental constraints. 
With a common factor of  $-i g m_f/2m_{W}$ being suppressed,
the couplings of the lighter and heavier CP-even Higgs bosons $h$ and $H$ 
to the tau, bottom and top quarks are, respectively, given by
\[
  \begin{tabular}{cccc}
   &  $\tau^- \tau^+$ &  $b \bar b$ & $t \bar t$ \\
$h$: \quad & $\; {- \sin\alpha/\cos\beta}\;\;  $ & $\; \; 
{- \sin\alpha/\cos\beta}
\;\;  $ & $\; \; {\cos\alpha/\sin\beta} \;\;  $\\
$H$: \quad & $ {\cos\alpha/\cos\beta}$ &  $ {\cos\alpha/\cos\beta}$ &  
${\sin\alpha/\sin\beta}$ 
  \end{tabular}
\]
Other relevant couplings of $h$ and $H$ are those to the weak gauge bosons, 
which have a factor of $\sin(\beta -\alpha)$ and 
$\cos(\beta -\alpha)$ relative to the SM values, respectively.

Although $\sin\alpha \approx 0$ is chosen in Eq.~(\ref{parameters}),
we do not consider the case when $\sin\alpha$ is fine-tuned to a
value very close to zero, say smaller than $10^{-3}$. In such a case,
the factor $-\sin\alpha/\cos\beta$ remains suppressed even if
$\tan\beta$ is extremely large. 
In the case when the smallest value of
$\sin\alpha$ is around $10^{-2}$ and $\tan\beta$ is larger than 20,
the factor $-\sin\alpha/\cos\beta$ becomes substantial and so the
lighter Higgs boson $h \to \tau^+\tau^-$ is no longer suppressed
\cite{cw}. 
Nevertheless, if $\sin\alpha$ is fine-tuned to $10^{-3}$, then
$\tan\beta$ can be much larger than 20, and even around 60 the decay
$h \to \tau^+\tau^-$ is still strongly suppressed.
On the other hand, if $\tan\beta < 10 $ 
the factor $\cos(\beta-\alpha)$ is not suppressed enough such that 
the heavier Higgs boson $H\to ZZ^{(*)} \to 4 \ell $ for $m_H = 130 - 600$ GeV 
might appear in the data.
In this work, we use a small $\sin\alpha \sim 10^{-2}$ and
$\tan\beta = 10-20$. 

\section{Experimental constraints}
The first constraint we must consider is the radiative decay of $B$ meson, 
$b \to
s\gamma$, as well as the $B$-$\overline{B}$ mixing, both of which receive
contributions from the charged Higgs boson. As shown in \cite{otto}, we are safe
if we take $m_{H^\pm} \ge 500$ GeV for $\tan\beta > 5$.
Yet, the most important experimental constraint on the 2HDM comes from the
$\rho$ parameter, which is theoretically and experimentally in the vicinity of
$1$, so that there is almost no room for deviations and it implies 
a strong restriction on the splitting of the Higgs boson masses.
In 2HDM $\Delta \rho$ receives contributions from all Higgs bosons given by
\cite{otto}
\begin{eqnarray}
\label{rho}
\Delta \rho^{\mbox{\tiny 2HDM}} &=& 
\frac{\alpha_{\rm em}}{4\pi \sin^2\! \theta_{\! \scriptscriptstyle W}
 M_{\! \scriptscriptstyle W}^2} \, 
\Big[ \, F( m_{\! \scriptscriptstyle A}, m_{\! \scriptscriptstyle {H^+}} ) 
 + \cos^2\!(\beta -\alpha) \, [ F(m_{\! \scriptscriptstyle {H^+}},m_h) 
- F(m_{\! \scriptscriptstyle A}, m_h) ] 
 \nonumber \\ && 
 + \sin^2\!(\beta -\alpha) \, [ F(m_{\! \scriptscriptstyle {H^+}}, m_{\! \scriptscriptstyle {H}} ) 
 - F(m_{\! \scriptscriptstyle A},m_{\! \scriptscriptstyle {H}})] \,\Big ]
 \nonumber \\
&& + \cos^2\!(\beta-\alpha) \Delta \rho^{\mbox{\tiny SM}}\!(m_{\! \scriptscriptstyle {H}}) 
   + \sin^2\!(\beta-\alpha) \Delta \rho^{\mbox{\tiny SM}}\!(m_h) \;,
\end{eqnarray}
where
\begin{eqnarray}
F(x,y) &=& {1\over 8}x^2 +{1\over 8}y^2-{1\over 4}\frac{x^2 y^2}{x^2-y^2} 
\log\!\left({x^2\over y^2}\right) \; ,
\nonumber \\
\Delta \rho^{\mbox{\tiny SM}}\!(M) &=& -
\frac{\alpha_{\rm em}}{4\pi \sin^2\! \theta_{\! \scriptscriptstyle W}
 M_{\! \scriptscriptstyle W}^2} \, \left[   3 F(M, M_{\! \scriptscriptstyle W})
- 3 F(M, M_{\! \scriptscriptstyle Z} )  +{1\over 2} (M_{\! \scriptscriptstyle Z}^2
-M_{\! \scriptscriptstyle W}^2) \right ] \;.
\end{eqnarray}
The $\Delta \rho^{\mbox{\tiny SM}}\!(M)$ is rather small and can be ignored, 
but the other terms are sizable if the mass difference
between the two mass arguments in $F(M_1, M_2)$ is large.  Just the first
term $F( m_{\! \scriptscriptstyle A}, m_{\! \scriptscriptstyle {H^+}} ) $ 
in Eq.(\ref{rho}) can be
of order $0.01$ for $m_A < 100$ GeV and $m_{H^\pm} = 500$ GeV. One simple
solution is to choose $m_A \approx m_{H^\pm}$ and $\cos(\beta - \alpha)
\approx 0$, then the 2HDM contributions to the $\Delta \rho$ become minuscule.

Unless $\tan\beta$ is of order $100$ \cite{otto},
for Higgs boson masses larger than 100 GeV the contributions to the
muon anomalous magnetic dipole moment is very small. 
We will ignore this constraint.

The next constraint is the current LHC data on the observed Higgs boson, 
defined by the signal strength parameter which is the production rate 
relative to the corresponding SM one:
\begin{equation}
R^{h}_{X} \equiv \frac{\sigma(pp \to h ) \times B(h \to X) }
                     {\sigma(pp \to h_{\rm SM} ) \times B(h_{\rm SM} \to X) }
\end{equation}
for $X = WW^*, ZZ^*, \gamma\gamma, \tau^+ \tau^-$.
Here we take the observed boson to be the lighter CP-even Higgs $h$.
\footnote{
There are a number of works in 2HDM or supersymmetry frameworks that take the
heavier CP-even Higgs boson or the pseudoscalar boson as the observed particle 
\cite{2hdm-heavy,susy}; or both CP-even Higgs bosons are nearly degenerate
\cite{jack}. We do not pursue these possibilities here.
}

The current values for $R^h_{WW}$ and $R^h_{ZZ}$ are consistent with the
SM and $R^h_{\gamma\gamma}$ is larger than 1, but $R^h_{\tau^+\tau^-}$ appears
to be suppressed. Here we do not mean to perform a full scan of the
parameter space, but only to emphasis that the parameter space of 
Eq.(\ref{parameters})
would give an enhanced diphoton rate $R^h_{\gamma\gamma}$ for the lighter
CP-even Higgs boson while keeping the $R^h_{WW}$ and $R^h_{ZZ}$ intact with
the SM values \cite{ours,sher}.  The bonus is the suppressed 
$R^h_{\tau^+\tau^-}$, which appears
to be consistent with both the CMS and ATLAS  experiments that 
have reached the tau-pair detection sensitivity limit.

\section{The heavier CP-even Higgs boson}
\begin{figure}[th!]
\includegraphics[width=4in]{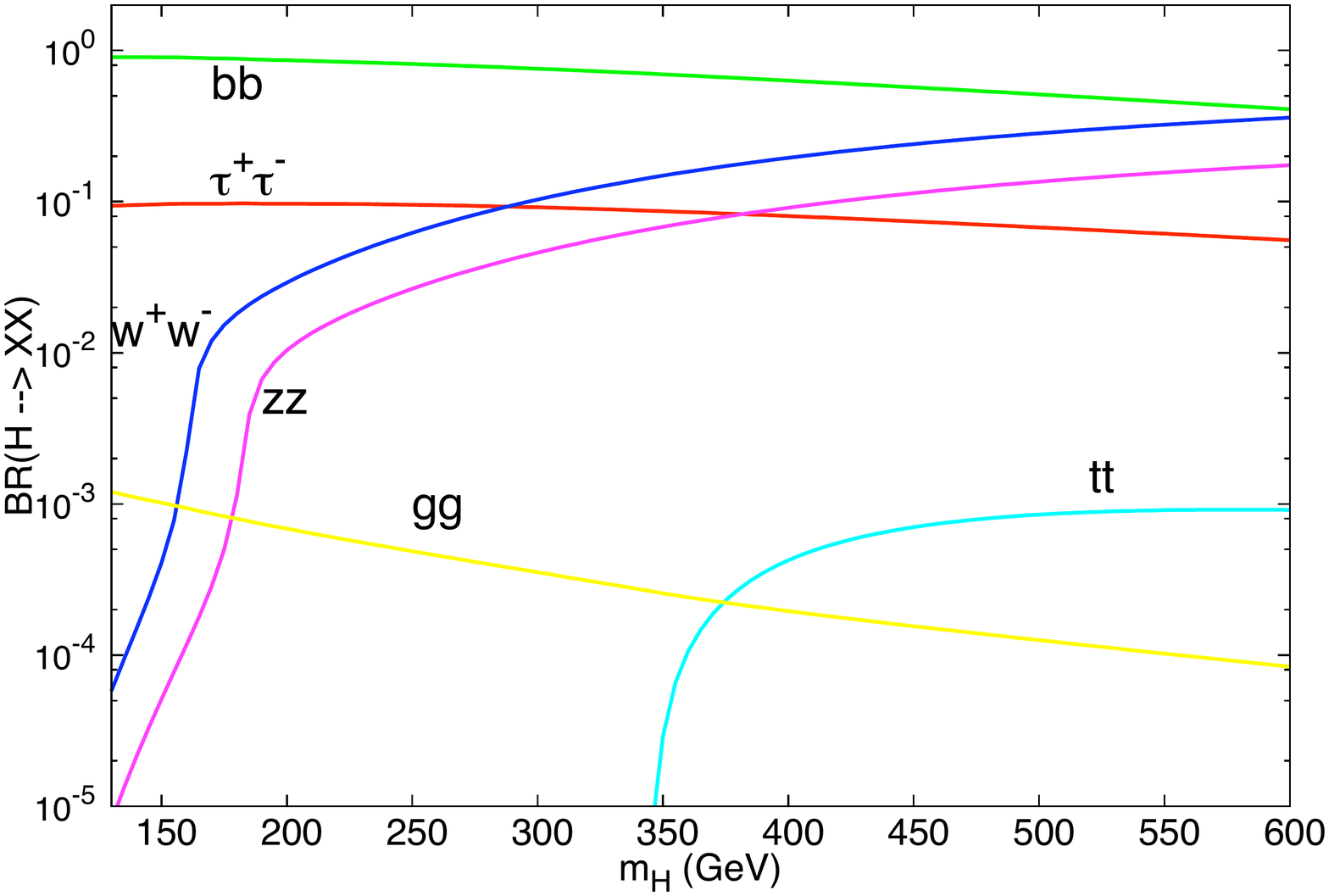}
\includegraphics[width=4in]{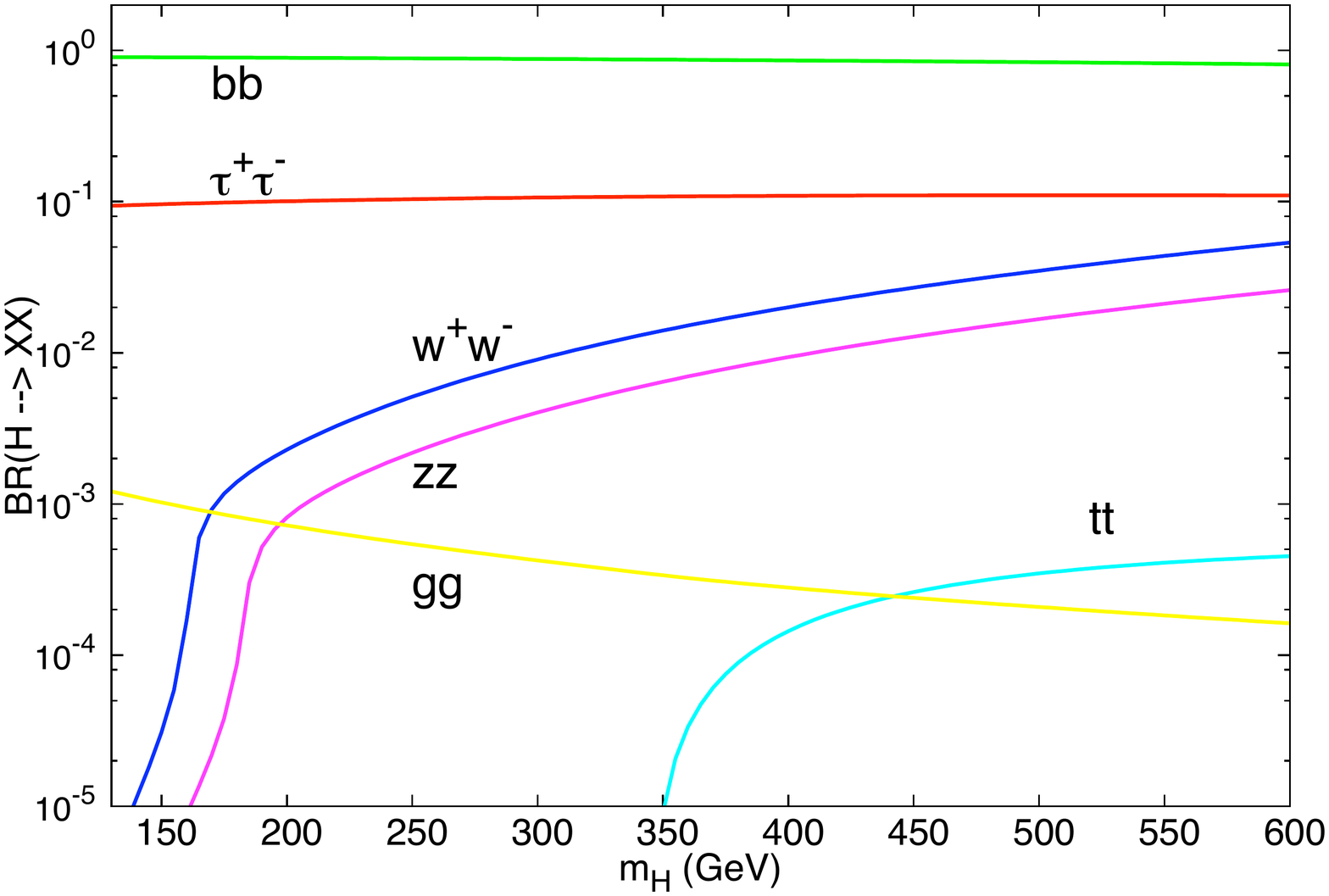}
\caption{\small \label{br}
Decay branching ratios for the heavier CP-even Higgs boson $H$ versus its mass $m_H$.
The parameters for the 2HDM are chosen as $\sin\alpha = 0.01$ and $\tan\beta$
=10 ({\it top}) and 20 ({\it bottom}).
}
\end{figure}

The scenario defined by the parameter space of Eq.(\ref{parameters}) 
would give an interesting signature for the heavier CP-even Higgs boson.  
Since $\cos(\beta - \alpha) \approx 1/\tan\beta$ for a large 
enough $\tan\beta$ and $\sin\alpha \approx 0$, the heavier CP-even
Higgs $H$ couples very mildly to $WW$ and $ZZ$ 
while the decays into $\tau^+\tau^-$ and $b\bar b$ are 
enhanced by $\tan\beta$. Therefore, 
the current constraint of $h_{\rm SM} \to ZZ^* \to 4\ell$ does not apply 
to $H$, so that its mass $m_H$ can be anywhere between $130$ GeV and 1 TeV.
\footnote{Similar idea was considered in Ref.~\cite{2higgs}.}
We show the decay branching ratios of 
$H$ versus $m_H$ in Fig.~\ref{br}, where we have used $\sin\alpha = 0.01$ and 
$\tan\beta$ = 10 ({\it top}) and 20 ({\it bottom}). Note that both 
the $\gamma\gamma$ and $Z\gamma$ branching ratios
are below $10^{-5}$ in this case and not shown. The $\tau^+ \tau^-$ 
branching ratio
varies from 5\% to 10\%, and the $WW^{(*)}$ and $ZZ^{(*)}$ are 
sub-leading as long as
$\tan\beta$ is large enough.
Note that if $\tan\beta < 10$, the $H\to ZZ^{(*)} \to 4\ell$ mode is 
not small enough such that
the current data can constrain a part of the mass range.

The production of $H$ goes through gluon fusion and the associated production
with a $b\bar b$ pair.  The production cross section of $H$ via gluon
fusion would be substantially smaller than that of the SM Higgs boson. 
This is because in the region of Eq.(\ref{parameters}) the top Yukawa 
is suppressed by
$m_t \sin\alpha/\sin\beta$ whereas the bottom Yukawa $m_b \cos\alpha/\cos\beta$
does not receive a large enough enhancement for $\tan\beta = 10-20$ to
make up for the mass suppression factor from $m_b$. 
Therefore, the gluon fusion cross section
would be at least a factor of 4 smaller than that of the SM Higgs boson.
The channel $gg \to H \to \tau^+ \tau^-$ would be buried 
under the current
$\tau^+\tau^-$ data.

\begin{figure}[th!]
\includegraphics[width=5in]{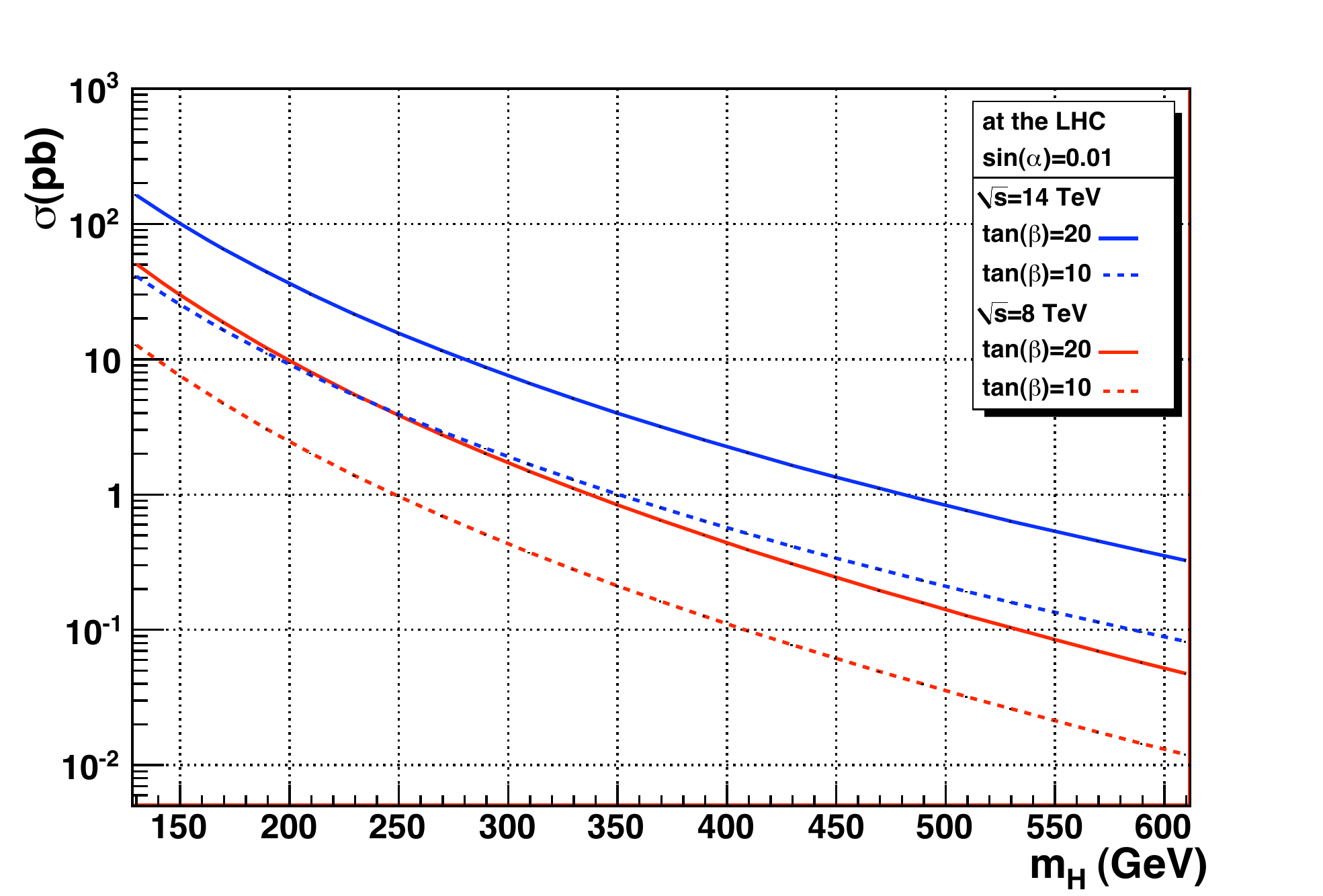}
\caption{\small \label{prod}
Production cross sections for $pp\to b\bar b H$ at the 8 TeV and 14 TeV LHC.
The parameters for the 2HDM are chosen as $\sin\alpha = 0.01$ and $\tan\beta$
=10 ({\it dashed}) and 20 ({\it solid}).
}
\end{figure}

On the other hand, the $pp\to b\bar b H$ receives a large $\tan\beta$ 
enhancement. 
We show the production cross sections for $pp \to b\bar b H$
at the 8 and 14 TeV LHC for $\sin\alpha = 0.01$ and $\tan\beta = 10,20$
in Fig.~\ref{prod}. The cross section can reach a level of $10-100$ pb
for $m_H = 130$ GeV.

\section{Signal-background analysis} 
In the following, we present the signal and backgrounds 
of $b\bar b \tau^+\tau^-$
final state in detector-simulation level by employing the 
{\sf Delphes} package \cite{delphes} inside 
{\sf MADGRAPH} \cite{mad}, with the most general settings for the LHC.

Dominant backgrounds come from 
$Z b \bar b$ and $Z jj$ 
(when the light-flavored quark jet or gluon jet is misidentified 
as a $b$ quark) production followed by $Z \to \tau^+\tau^-$.
We employ the $B$-tagging with a constant efficiency of 60\% and 
the mistag efficiency from a light-flavored quark or a gluon to be 1\%
\cite{ferro}. The selection cuts are
\begin{eqnarray}
&&E_{T_{j,b}} > 30 \;{\rm GeV},\;\; |\eta_{j,b}| < 2.5, \;\; \Delta R_{jj} > 0.5
\nonumber \\
&&p_{T_\tau} > 30\;{\rm GeV}, \;\; |\eta_\tau| < 2.5,\;\; \Delta R_{\tau j} > 0.5
\label{cuts}
\end{eqnarray}
We calculate the differential cross sections under these cuts
and $B$-tagging efficiency, and use the Delphes for detector simulation
and reconstruction of the tau leptons, $b$-jets, and the 
Higgs boson and $Z$ boson peaks.
\footnote
{The $p_T$ reconstruction requirements for the tau leptons coming from the 
Higgs decay and from the background are treated differently, in order
to correctly reconstruct the mass peak of the Higgs boson and the $Z$ boson.
We rescale the distributions to their corresponding cross sections before
adjusting the $p_T$.
}

\begin{figure}[th!]
\includegraphics[width=5.5in]{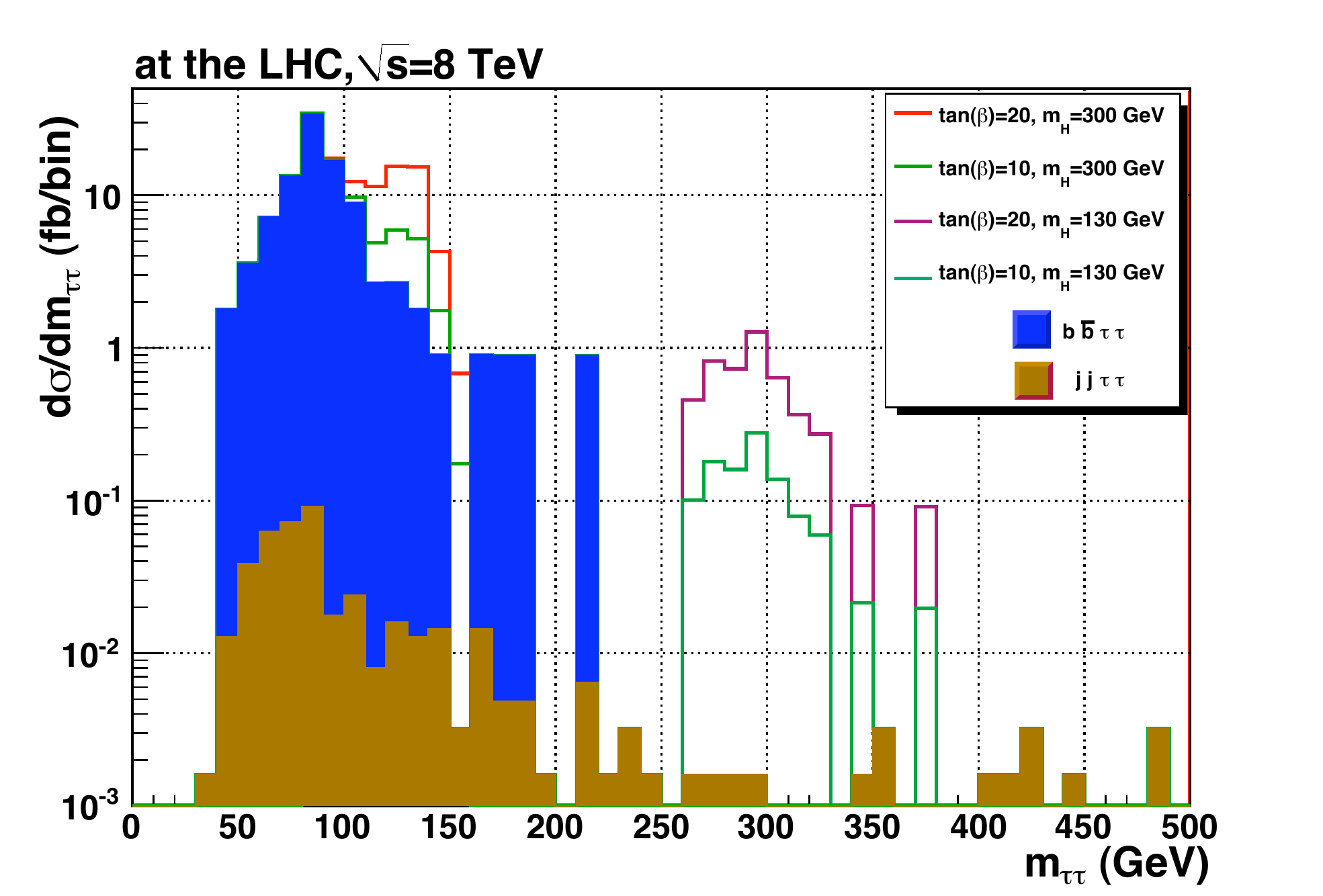}
\caption{\small \label{signal}
Differential cross section $d\sigma/dM_{\tau^+\tau^-}$
versus $M_{\tau^+\tau^-}$ for the continuum background coming from 
$b\bar b \tau^+\tau^-$ and $jj\tau^+\tau^-$ and for the hypothetical signals of 
the heavier CP-even Higgs boson $H$ of 130 and 300 GeV. 
Details of detector simulations and normalization are described in the text.
}
\end{figure}

We show the continuum background for the sum of $b\bar b \tau^+\tau^-$ 
and $jj\tau^+\tau^-$ versus $M_{\tau^+\tau^-}$ in Fig.~\ref{signal}.
The majority of the background comes from the $Z$ boson with a peak around
$m_Z$ in the figure. 
Hypothetical signals of $m_H = 130$ GeV and 300 GeV are added to
the background in the figure.
The $jj\tau^+\tau^-$ background decreases to a negligible
level while the dominant background comes from $b\bar b \tau^+\tau^-$. However, 
the signal for
$\tan\beta = 10$ stands somewhat above the background and the signal
for $\tan\beta = 20$ is unambiguously discernible.
The cross sections under the 130 GeV and 300 GeV Higgs boson $H$ at the 8 TeV
LHC are $10$ and $1$ fb, respectively, for $\tan\beta = 10$, and about a 
factor of $4$ larger for $\tan\beta =20$. The cross section under the 
Higgs boson peak should be large
enough for observation.

\section{Discussion} 

We offer a few comments as follows.
\begin{enumerate}
\item The process $pp \to t\bar t \to b W^+ \bar b W^- \to 
b \tau^+ \nu_\tau \bar b \tau^- \bar \nu_\tau$ would also give a 
$b\bar b \tau^+\tau^-$ final state, but it can be
reduced to a negligible level by requiring no $p_T$ missing.

\item The mass range hinted by the $b\bar b$ excess at the Tevatron 
\cite{tevatron} might be explained by the heavy CP-even Higgs boson $H$ in
this scenario with a wide mass range peaked at around 135 GeV. 

\item In this work, the $\tau$ identification and reconstruction 
are taken into account in
the Delphes, which has an efficiency similar to actual experiments.
The tau efficiency obtained by experiments 
is around 50\% - 60\% with the fake rate from jets around 1\% \cite{cms-tau}.

\item  The process $pp \to b \bar b H \to b \bar b b \bar b$ is also interesting
and expected 
to be enhanced in the model. However the severe QCD continuum background 
and the combinatorics must also be scrutinized.

\item If the $\tau^+\tau^-$ and $b\bar b$ modes continue to be suppressed
in the observed 125 GeV Higgs boson, there must be another Higgs-like boson
that couples to $b$ and $\tau$. Therefore, $pp\to b\bar bH$ production
followed by $H\to \tau^+ \tau^-$ in the $b\bar b \tau^+\tau^-$ final state
is a rather general channel to probe this scenario. In particular, the
$b$ coupling in 2HDM is enhanced by $\tan\beta$ such that the $H$ 
can be observable at the LHC.
\end{enumerate}

We have shown that if the observed Higgs boson continues to give 
an enhanced $\gamma\gamma$ rate but a suppressed $\tau^+\tau^-$ rate, it 
is likely that there exists another heavier CP-even Higgs boson decaying into
$\tau^+\tau^-$.  We showed in the 2HDM 
that using the associated production $b\bar b H$ followed
by $H\to \tau^+\tau^-$ we can detect this heavier CP-even Higgs boson at
the LHC. It is of the highest priority that the LHC experiments now search
for the $b\bar b \tau^+\tau^-$ final state to uncover another Higgs boson.

\section*{Acknowledgment}  
This work was supported the National Science
Council of Taiwan under Grants No. 99-2112-M-007-005-MY3 and No.
101-2112-M-001-005-MY3 as well as the WCU program through the KOSEF
funded by the MEST (R31-2008-000-10057-0).  TCY would like to
acknowledge the hospitality from NCTS where progress of this work was
partly made.


\end{document}